\documentclass[11pt]{cernrep}

\usepackage{graphicx}
\usepackage[reqno]{amsmath}
\usepackage{here}

\newcommand{\BBbar}{\ensuremath{\mathrm{B}\overline{\mathrm{B}}}}
\newcommand{\bmes}{\ensuremath{\mathrm{B}}}
\newcommand{\bmin}{\ensuremath{\mathrm{B}^-}}
\newcommand{\bplus}{\ensuremath{\mathrm{B}^+}}
\newcommand{\bz}{\ensuremath{\mathrm{B}^0}}
\newcommand{\bzbar}{\ensuremath{\overline{\mathrm{B}}^{\:0}}}
\newcommand{\ccbar}{\ensuremath{c\overline{c}}}
\newcommand{\dcalc}[1]{\ensuremath{\mathrm{d}#1}}
\newcommand{\dmes}{\ensuremath{\mathrm{D}}}
\newcommand{\dedx}{\ensuremath{\mathrm{d}E/\mathrm{d}x}}
\newcommand{\dstarplus}{\ensuremath{\mathrm{D}^\ast(2010)^+}}
\newcommand{\dz}{\ensuremath{\mathrm{D}^0}}
\newcommand{\dzbar}{\ensuremath{\overline{\mathrm{D}}^{\:0}}}
\newcommand{\gev}{\ensuremath{\mathrm{GeV}}}
\newcommand{\kl}{\ensuremath{\mathrm{K}_L^0}}
\newcommand{\kmes}{\ensuremath{\mathrm{K}}}
\newcommand{\kmin}{\ensuremath{\mathrm{K}^-}}
\newcommand{\kplus}{\ensuremath{\mathrm{K}^+}}
\newcommand{\ks}{\ensuremath{\mathrm{K}_S^0}}
\newcommand{\kz}{\ensuremath{\mathrm{K}^0}}
\newcommand{\ra}{\ensuremath{\rightarrow}}
\newcommand{\tesla}{\ensuremath{\mathrm{T}}}
\newcommand{\ufours}{\ensuremath{\Upsilon(4S)}}
\newcommand{\vp}{\ensuremath{\vec{p}}}
\newcommand{\ycp}{\ensuremath{y_{\text{CP}}}}

\newcommand{\electron}{\ensuremath{\mathrm{e}^-}}
\newcommand{\positron}{\ensuremath{\mathrm{e}^+}}
\newcommand{\epem}{\positron \electron}

\def\smis{S_{tail}}
\def\smisbg{S_{tail}^{BG}}
\def\sbg{S_{BG}}
\def\fmis{f_{tail}}
\def\fmisbg{f_{tail}^{BG}}
\def\ftbg{f_{\tau_{BG}}}
\def\tsig{\tau_{SIG}}
\def\tbg{\tau_{BG}}

\begin{document}

\title{STATISTICAL PRACTICE AT THE BELLE EXPERIMENT, AND SOME QUESTIONS}

\author{Bruce Yabsley}

\institute{Virginia Polytechnic Institute and State University, Blacksburg VA}

\maketitle

\begin{abstract}
The Belle collaboration operates a general-purpose detector at the KEKB
asymmetric energy $e^+ e^-$ collider, performing a wide range of measurements in
beauty, charm, tau and 2-photon physics. In this paper, the treatment of
statistical problems in past and present Belle measurements is reviewed. 
Some open questions,
such as the preferred method for quoting rare decay results,
and the statistical treatment of the new $\bz/\bzbar \to \pi^+\pi^-$ analysis,
are discussed.
\end{abstract}

%
%

\section{INTRODUCTION}
\label{section-intro}

My ambitions for this conference are to recover my luggage, which went missing
four days ago, and to find answers to some questions about statistical practice
at Belle.\footnote{My luggage arrived on the Thursday morning, about 30 hours 
	before I left Durham. Some of my questions were answered, at least in
	part. More of this anon.}
I suspect I'm not the only one with an agenda in this area.
From the point of view of the Belle spokesmen, it would be far better if 
I could articulate workable guidelines for our use of statistical methods
\ldots rather than finding fault with our present practice case-by-case.
As I hope to show, it is much easier to criticise our statistical practice than
it is to suggest alternatives,
although I have made some tentative steps in that direction. 

Analyses at Belle are not all of the same kind,
and the ``statistical environment'' varies from one study to another.
After reviewing the experiment itself (section~\ref{section-belle})
and some general issues (section~\ref{section-practice}), 
I hope to give you a feeling for the main types of analysis, and the 
statistical issues in each case (section~\ref{section-analyses}). 
Of particular interest are the searches for rare \bmes-meson decays
(section~\ref{subsec-analyses-rare}), where there is a tradeoff between
``purist'' statistical concerns and important practical issues; and our new
analysis of  $\bz/\bzbar \to \pi^+\pi^-$ decays (section~\ref{section-pipi}).
Interpretation of this result is surprisingly difficult,
due to the unusual configuration of the parameter space,
as well as some features of the analysis. The first Belle paper on this topic
is two weeks from journal submission, so you may have the chance to influence
our statistical practice in real time!

%
%

\section{THE BELLE EXPERIMENT}
\label{section-belle}

\begin{figure}
  \begin{center}
    \includegraphics{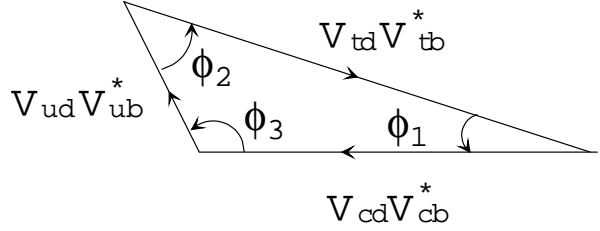}
  \end{center}
  \caption{The unitarity triangle relevant to \bmes\ decays.
	In all unitarity triangles, the angles differ from $0^\circ$ and
	$180^\circ$ if the KM phase is nonvanishing.
	In \emph{this} triangle, the angles are expected to be far from
	$0^\circ$ and $180^\circ$, leading to readily measurable effects.} 
  \label{figure-unitarity}
\end{figure}

The main ``point'' of Belle is to test the Kobayashi-Maskawa model of 
CP-violation~\cite{km}, in which the phenomenon is entirely due to the
irreducible complex phase of the quark mixing matrix.
The simplicity of this model makes it highly predictive. 
In the absence of extra generations or isosinglet quarks, we have the familiar
$3 \times 3$ matrix
\[
  \begin{pmatrix}
	V_{ud}	&  V_{us}	&  V_{ub}	\\
	V_{cd}	&  V_{cs}	&  V_{cb}	\\
	V_{td}	&  V_{ts}	&  V_{tb}
  \end{pmatrix}
\]
(the so-called CKM matrix), and unitarity gives a number of relations of the
form $V_{td} V_{tb}^\ast + V_{cd} V_{cb}^\ast + V_{ud} V_{ub}^\ast = 0$. 
This particular ``unitarity triangle'' (see Fig.~\ref{figure-unitarity})
is especially interesting: all its interior angles $\phi_{1,2,3}$ are believed
to be far from $0^\circ$ and $180^\circ$,
and may be measured using time-dependent asymmetries between \bz\ and \bzbar\
decays to appropriate states. 
In other words: CP-violating asymmetries in these decays are expected to be
large. The recent measurement of $\sin 2\phi_1$ by Belle~\cite{belle-sin2phi1}
confirms that this is true, at least regarding $\phi_1$: based on a study of
$\bz,\bzbar \to J/\psi\,\ks$ and related decays,
we found $\sin2\phi_1 = 0.99 \pm 0.14 \pm 0.06$.
(This is the first observation of a CP-violating effect outside the neutral
kaon system. In a spirit of friendliness let me also cite the similar, and
essentially simultaneous, result from BaBar~\cite{babar-sin2phi1}.)
The $\bz/\bzbar \to \pi^+\pi^-$ analysis discussed in section~\ref{section-pipi}
is sensitive to the angle $\phi_2$.

The Belle detector~\cite{belle-detector} is a classic solenoidal tracking
detector ($|\vec{B}| = 1.5\,\tesla$) with some extra features. 
Most notable is its asymmetry,
to optimize acceptance of events with a centre-of-mass boosted
by $\beta\gamma = 0.425$ in the lab frame. (The storage rings of KEKB collide 
$8\,\gev\;e^-$ on $3.5\,\gev\;e^+$, \emph{i.e.}\ $\sqrt{s} = 10.58\,\gev$,
yielding $\ufours \to \bmes\bar{\bmes}$ decays or nonresonant $q\bar{q}$,
$\tau^+\tau^-$ \emph{etc.})
The drift of the $\bz/\bzbar$ prior to decay is measured using 
a silicon vertex detector, allowing measurement of time-dependent
asymmetries (see section~\ref{subsec-analyses-timing}).
Particle ID over the full momentum range for \bmes-daughters is obtained
using aerogel \v{C}erenkov counters as well as the more traditional
\dedx\ and time-of-flight measurements: this improves \bmes\ flavour-tagging, 
and allows identification of rare \bmes\ decays where we hope to find evidence
of ``direct'' CP violation (section~\ref{subsec-analyses-rare}).
The flux return is instrumented with RPCs to identify $\mu^\pm$
(as matched tracks),
improving the efficiency and purity of $J/\psi \to \mu^+\mu^-$ reconstruction,
and \kl\ (as neutral clusters), allowing measurement of $\sin2\phi_1$ using
$J/\psi\,\kl$ and related modes.  

The collaboration itself is of order 250 physicists,
from over 50 institutes in 12 countries.
Perhaps half of the analysis effort is directed towards headline studies
like $J/\psi\,\kmes^{(\ast)}$ and $\pi^+\pi^-$, with the rest thinly spread
over rare \bmes\ decays, CKM matrix elements 
(\emph{via} measurements of semileptonic decays),
charm studies, $\tau$ and $2\gamma$ physics. 
Most of this work proceeds through some mix of local ``specialties'' and
individual initiative, with fairly light coordination from the centre.
Which brings us to my next point.

%
%

\section{ABOUT STATISTICAL PRACTICE AT BELLE}
\label{section-practice}

\begin{quote}
  \emph{
    In those days there was no king in Israel;
    everyone did what was right in his own eyes. 
  }
  \\ \hspace*{\fill}
  Judges 21:25
\end{quote}

\noindent
The key to understanding statistical practice at Belle is that 
\emph{there is no official policy}.
Partly as a result of this, our practice is inconsistent.
The following is my own take on this state of affairs.
\begin{itemize}
  \item	This is not nearly as bad as it sounds:
	\vspace*{1ex}
	\begin{itemize}
	  \item	Most of our procedures \emph{are} motivated, either by 
		principle or tradition. In some places traditional methods have
		perhaps been mistaken for rigorous formulae, but this is a
		higher-order problem: one of education, not discipline.
	  \item	We do describe our procedures in our papers, and this is
		more important than consistency.
		The description is sometimes incomplete, but I hope we are
		becoming more sensitive to this.
	\end{itemize}
	\vspace*{1ex}
  \item	There \emph{are} things which should, and may change.
	I will mention some of them below.
  \item	In avoiding anarchy, we do not want to become an authoritarian state.
\end{itemize}
The last of these points is worth stressing. Most of us spend our time
going about our own business, and many of us have significant freedom to 
\emph{define} what that business is. In part this is our way of working:
250 physicists may sound like a lot, but the number of potential physics topics
is vast, and individuals ``fanning out'' opportunistically is a good way
to cover them. But it is a \emph{modus vivendi} as well.
We live with each others' egos, and satisfy our own, by spending
most of our time on our own affairs.

Prescriptive policies have a way of cutting across this, and we tend to avoid
them. Of course we set standards for papers (our \emph{ad hoc} paper refereeing
committees are almost the only committees worthy of the name) and areas of 
analysis are run by conveners, who act as facilitators and clearinghouses for
information; occasionally, we also commission work. But to write a recipe-book
for statistical practice, which
all published analyses were expected to follow---this would be unprecedented.
I don't think people would be happy with it, and I doubt ``the authorities'' 
would exert themselves to enforce it \ldots change, if it occurs, is more
likely to proceed \emph{via} the winning of hearts and minds.
I suspect consciousness-raising in analysis groups, and among the individuals 
who sit on those refereeing committees, is the key.\footnote{When I began
	preparing this study I thought a set of usable public tools was the
	key. While these have their place, and I haven't abandoned the ambition
	of providing some, I've come to believe that issues of principle, and
	some genuinely open questions, must be addressed first. The remainder
	of this talk hopefully shows why.}

%
%

\section{THE BELLE ANALYSES}
\label{section-analyses}

The analyses themselves, by which I mean analyses we have written up and 
published (or have under review), can be divided into four broad categories:
\begin{enumerate}
  \item	the flagship time-distribution fit analyses (usually concerning CPV);
  \item	searches for rare hadronic $B$ decays;
  \item	fits over a flattish background;
  \item	systematic-dominated analyses.
\end{enumerate}
This is a grouping according to statistical problems, rather than common work.
The second category corresponds roughly to one of our analysis groups,
which does lead to a certain uniformity of method. 
Otherwise, these categories cut across Belle's administrative divisions. 
I will consider them in turn, with the aid of an example in each case.

%
%

\subsection{The flagship time-distribution fits}
\label{subsec-analyses-timing}

The unitarity angle measurements mentioned above, based on 
$\bz/\bzbar \to J/\psi\,\kmes^{(\ast)}$ \emph{etc.}~\cite{belle-sin2phi1} and
$\pi^+\pi^-$~\cite{paper-pipi}, are performed by fitting
decay-time distributions.
To illustrate the method I have chosen a simpler analysis, which (for our
purpose) has the same features: a measurement of the $\dz-\dzbar$
mixing parameter \ycp, through the lifetime difference of neutral \dmes-mesons
decaying to $\kmin\kplus$ (a CP eigenstate) and $\kmin\pi^+$~\cite{paper-ycp}.

After selecting $\dz \to \kmin\kplus$ and $\kmin\pi^+$ decays\footnote{Here
	and in general, I imply the inclusion of charge-conjugate modes.}
(imposing particle ID and decay angle cuts, and a momentum cut to veto
\bmes-daughters)
we fit the tracks to a common vertex, and extrapolate the resulting
\dmes-momentum $\vp_{\dmes}$ to the interaction region:
this gives us the flight length and thus the decay time. 
Thanks to our good kaon/pion separation the \dmes-decay samples are fairly 
clean, so we avoid $\dstarplus \to \dz \pi^+$ tagging, keeping the samples
as large as possible. Of course some combinatorial $\kmin\kplus/\kmin\pi^+$  
is accepted, and we estimate the probability for each candidate $i$ to be a true
\dmes-decay using 
\begin{equation}
  f_{SIG}^i = \frac{N_{SIG}(M^i)}{\left(N_{SIG}(M^i) + N_{BKG}(M^i) \right)},
  \label{def-fsig}
\end{equation}
where $M^i$ is the mass of the candidate, and signal and background fractions
$N_{SIG}(M^i)$ and $N_{BKG}(M^i)$ are taken from a fit to the 
mass distribution of all candidates. The distributions are uncomplicated, and
double-Gaussian fits over linear backgrounds are sufficient for the purpose.
We accept events well into the tail---out to $6\sigma$ in the mass---for 
reasons that should become clear.

We then perform an unbinned maximum likelihood fit to the events,
using the function
\begin{eqnarray}
&&\mathcal{L}(	\tsig, S, \smis, \fmis,
		\tbg, \ftbg,
		\sbg, \smisbg, \fmisbg
) \nonumber \\
&&\quad= \prod_i \left[f_{SIG}^i\int^\infty_0 dt^\prime
\frac{1}{\tau_{SIG}}e^{\frac{-t^\prime}{\tau_{SIG}}}
R(t^i-t^\prime;\sigma_t^i, S,\smis,\fmis )\right.
\label{def-dtime-lhd}
\\
&&\quad\quad\left.+
(1-f_{SIG}^i)\int^\infty_0 dt^\prime
\{ f_{\tau_{BG}}\frac{1}{\tau_{BG}}e^{\frac{-t^\prime}{\tau_{BG}}} 
+(1-f_{\tau_{BG}})\delta(t^\prime)\}
R(t^i-t^\prime;\sigma_t^i, \sbg,\smisbg,\fmisbg )\right] \nonumber
\end{eqnarray}
which we also use to terrify small children. It is much less complicated than
it looks. The first and second lines are the signal and background parts
respectively: the underlying time-distribution of the signal is exponential,
while that of the background is modelled by a fraction ($f_{\tau_{BG}}$) with
lifetime (\emph{e.g.}\ charm daughters),
following $e^{-t^\prime/\tau_{BG}}$, and a fraction without,
following $\delta(t^\prime)$; $R$ is a double-Gaussian resolution function. 
The nine parameters shown are floated in the fit, so the background properties
are fitted along with the signal: this is the reason for including the region
$3\sigma < |M^i - m_{\dmes}| < 6\sigma$ in the fit, providing a background-rich
sample which largely determines the background parameters.

(In fact there are eighteen fitted quantities,
because there are two functions~(\ref{def-dtime-lhd}): one each for $\kmin\pi^+$
and $\kmin\kplus$ decays. Our fit maximises the grand likelihood
$\mathcal{L} = \mathcal{L}_{\kmes\pi} \cdot \mathcal{L}_{\kmes\kmes}$,
replacing the $\kmin\kplus$ lifetime by 
$\tau_{\kmes\kmes} = \tau_{\kmes\pi} / (1 + \ycp)$.
We find a null result, by the way: $\left.\ycp = (-0.5\pm1.0\pm0.8)\%.\right)$

If you look closely at the resolution term, you'll notice that the proper-time
error for each event is given by $\sigma_t^i$: an event-dependent quantity.
Due to variations in track and \dmes-vertex quality, the estimated vertexing
and therefore proper-time errors vary from event to event,
by a factor of a few. (Kinematic variations also play a role.) 
We scale these errors by global factors $S$ (for the core Gaussian; close to 1)
and $\smis$ (for the tail Gaussian; $> 1$), but the full function $R$ varies
event-to-event, as does the signal fraction $f_{SIG}^i$.
Any binned fit to the data would therefore need to have multidimensional
bins---\emph{many}-dimensional,
for a complicated analysis like $\sin2\phi_1$---and to avoid this, we resort to unbinned fits.

And so to the statistical issues raised by the timing-distribution fits:
\begin{enumerate}
  \item	\emph{How do we estimate goodness-of-fit to our timing distributions?}
	While standard measures exist for binned fits, 
	there is no accepted goodness-of-fit test for unbinned maximum 
	likelihood. Some effort has recently gone into finding a method
	\ldots if indeed it's possible~\cite{kayk}.
	In the meantime the lack of such a
	method is a nuisance, since we have nontrivial resolution functions
	which we fit from the data. How would we know if the functional form
	were wrong; and would it matter?
	\vspace*{1ex}
	\begin{enumerate}
	  \item	In the case of \ycp, we perform extensive systematic checks
		by varying cuts, signal-to-background ratios and the like;
		and trace some biases to their origin by turning effects on
		and off in our detector Monte Carlo. This doesn't so much
		assure us that the fit is good, as that any variations of the 
		fit, or problems with it, have a controlled effect on \ycp.
	  \item	For $\sin2\phi_1$, we test the (\emph{very} complicated)
		resolution function
		by using it in the measurement of \bmes-lifetime: a simpler,
		and much-higher-statistics task, than the asymmetry fit. 
		We check for biasses by fitting null-asymmetry samples which
		are similar to our signal: $\bplus \to J/\psi\,\kplus$,
		for example. And we compare our results to  ensembles of fits
		to toy Monte Carlo datasets \ldots although there may be less
		information in this last check than we once
		thought~\cite{kayk}.
	\end{enumerate}
	\vspace*{1ex}
	This is all good and necessary, and helps us (and our journal referees)
	to sleep at night. But if a decent goodness-of-fit test existed, we
	would obviously want to use that \emph{as well}, and I for one would
	be glad if someone developed such a thing. Or could convince me not to
	worry about it.
  \item	\emph{How should we combine our errors?}
	This is a problem we have largely postponed,
	as our unitarity angle analyses are still statistically limited.
	For \ycp, statistical and systematic errors are of equal magnitude,
	and we estimate the total error $\Delta$ using the familiar
	$\Delta^2 = \sigma^2\,\text{(stat.)}\, + \delta^2\,\text{(syst.)}$.
	Familiar is, of course, not the same as ``correct''.
  \item	\emph{How should we estimate confidence intervals?}
	For \ycp, what we \emph{do} is to treat the $\Delta$ just defined as a
	Gaussian-distributed error. (Yes, I know that's an assumption.)
	For the unitarity angle analyses, confidence intervals \ldots
	are a can of worms.  See section~\ref{section-pipi}
\end{enumerate}

%
%

\subsection{Searches for rare hadronic $B$ decays}
\label{subsec-analyses-rare}

Rare decay analyses are simpler, at least on the surface.
These studies are motivated by the search for ``direct'' CP violation,
\emph{i.e.}\ CP asymmetry of decay amplitudes. 
Decays which are Cabibbo, CKM or colour-suppressed, or proceed via loop 
diagrams, are a good place to look for direct CPV: mechanisms with
different CKM structure (such as Penguins and $V_{ub}$ tree diagrams)
can compete, with similar magnitudes; interference can lead to CP violation.
For similar reasons, 
``New Physics'' (non-Standard Model effects) can be expected to contribute,
since the competing Standard Model processes are suppressed.

There are many (possible) rare decay modes, but the analyses tend to follow the
pattern established in~\cite{paper-hh}.
As an example, let's take the somewhat simpler publication on
$\bmes \to \eta' \kmes$ and $\eta' \pi$ decays~\cite{paper-etapk}.

\subsubsection{What we do: the $\bmes \to \eta' \kmes\; (\eta' \pi)$ analysis}
\label{subsub-etapk}

\begin{figure}[H]
  \begin{center}
    \includegraphics[bb=0 0 566 520,height=6cm,width=12cm]{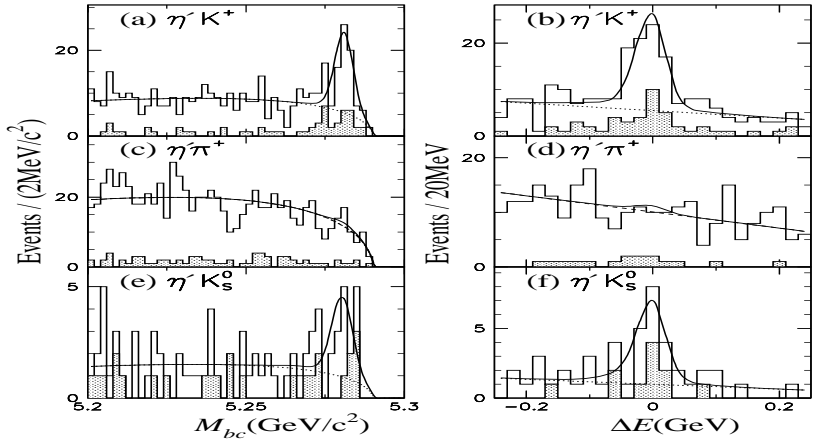}
  \end{center}
  \caption{From the $\eta' h$ paper~\cite{paper-etapk}:
	$M_{bc}$ and $\Delta E$ projections for $\eta^{\prime}\kplus$,
	$\eta^{\prime}\pi^+$ and $\eta^{\prime}K^0_S$ events.
	The shaded histograms show $\eta^{\prime}\to \eta \pi^+\pi^-$ events 
	while the open histograms show $\eta^{\prime}\to \eta \pi^+\pi^-$
	and $\eta^{\prime}\to \rho\gamma$ combined.
	The superimposed curves show the fits to $M_{bc}$ and $\Delta E$
	(solid) and the background component in the fit (dashed). 
	Note the negligible $\eta^{\prime}\pi^+$ yield.
	}
  \label{figure-etapk-fig3}
\end{figure}

After making an event selection, continuum (\emph{i.e.}\ $\epem \to q\bar{q}$)
backgrounds are suppressed using a likelihood ratio formed from
(i) a Fisher discriminant of event shape variables,
(ii) the production angle of the \bmes-candidate, and
(iii) for $\eta' \to \rho\gamma$ decays, a helicity variable.
A cut on the likelihood ratio gets rid of 70--90\% of the background
while keeping most of the signal.
The signal is then isolated in $M_{bc}$ and $\Delta E$ (beam-constrained mass
and energy-difference), which exploit the 
constrained kinematics of $\epem \to \ufours \to \BBbar$ events: the results
are shown in Fig.~\ref{figure-etapk-fig3} for 
$\bplus \to \eta' \kplus,\; \eta' \pi^+$ and $\bz \ra \eta' \kz$.
Signal events appear in Gaussian peaks at $M_{bc} = 5.28\,\gev/c^2$
and $\Delta E = 0$; continuum backgrounds follow a phase-space-like function 
due to ARGUS~\cite{argus-function} in $M_{bc}$, and a linear form in $\Delta E$.
(Background shape parameters are set from appropriate sideband data,
and cross-checked in the Monte Carlo simulation.)

Note that $\eta' \kz$ is near the edge of our sensitivity, and that we
see no $\eta' \pi^+$ peak. We assess the significance of a yield using
$\Sigma \equiv \sqrt{-2\ln(\mathcal{L}_0 / \mathcal{L}_{max})}$,
where $\mathcal{L}_{max}$ is the maximum likelihood
returned by the fit (here, an unbinned fit to $(M_{bc},\Delta E)$),
and $\mathcal{L}_0$ is the likelihood at zero yield.
For the cases
\begin{description}
  \item[$\Sigma \geq 3 \equiv$ ``significant'':]
	we quote a \emph{central value}, but no confidence interval
	(\emph{e.g.}\ $\bz \to \eta' \kplus$);
  \item[$\Sigma < 3 \equiv$ ``not significant'':]
	we quote an \emph{upper limit}, but no central value
	(\emph{e.g.}\ $\bz \to \eta' \pi^+$).
\end{description}
There are some full-reporting issues here, but let's set them aside.
The \emph{upper limit} (usually at 90\% C.L.) is calculated using the 
notorious method of \emph{``integrating the likelihood function''},
followed by addition of one unit of systematic error!
(The results are shown in Table~\ref{table-etapk-tab2}.)
Where we measure CP-asymmetries between (say) \bplus\ and \bmin\ decays, 
intervals are constructed in the same spirit: for
$A_{CP}(\bmes^\pm \to \eta' \kmes^\pm) = x^{+\sigma^+}_{-\sigma^-}\,\text{(stat.)}\;^{+\delta^+}_{-\delta^-}\,\text{(syst.)}$, we set a 90\% C.L. interval
$(x - 1.64\sigma^- - \delta^-,\; x + 1.64\sigma^+ + \delta^+)$.

%
%
\begin{table}[t]
  \begin{center}
    \caption{
	From the $\eta' h$ paper~\cite{paper-etapk}:
	Branching fraction ($BF$) or 90\% C.L. limit, and significance
	($\Sigma$) for Belle, compared with 
	CLEO~\cite{cleo-etapk} and BABAR~\cite{babar-etapk} results,
	and theoretical expectations~\cite{theor,sanda}. 
	The branching fractions are in units of $10^{-6}$.}
    \label{table-etapk-tab2}
\begin{tabular}{lcrccc}\hline\hline   
Mode & This measurement($BF$) & $\Sigma$ 
     & CLEO  & BABAR &   Theory \\\hline
$B^+\to\eta^{\prime}K^+$ & $79^{+12}_{-11}\pm 9$ & 12.0 
                         & $80^{+10}_{-9}\pm 7$  & $62\pm 18 \pm 8 $ & 21--53 \\
$B^+\to\eta^{\prime}\pi^+$ & $<7$  & 0.0 
                           & $<12$ & - & 1--3 \\
$B^0\to\eta^{\prime}K^0$ & $55^{+19}_{-16}\pm 8$ & 5.4 
                         & $89^{+18}_{-16}\pm 9$ & $< 112$ & 20--50 \\
\hline\hline
\end{tabular}
  \end{center} 
\end{table}

\subsubsection{Why what we do is not so bad}
\label{subsub-flatprior}

Like many of you, I have little good to say about this technique.
A likelihood $\mathcal{L}(\mu ; x)$ for parameter(s) $\mu$ given 
measurement(s) $x$ is nothing other than the \emph{probability density}
$p(x ; \mu)$ to obtain the observed data, if the underlying parameter
really were $\mu$. It is thus a density \emph{in x}, and to ``integrate''
a density in the wrong variable is confused.
(Try this with a Gaussian
$p(x ; \mu) = (2\pi\sigma^2)^{-1/2} \exp(-(x-\mu)/2\sigma^2)$,
integrating $\int \dcalc{\mu}\, p(x ; \mu)$ for fixed $x$---as 
opposed to integrating it over $x$---and you'll see what I mean.)
And do not get me started on the addition of one unit of systematic error \ldots
at least until section~\ref{section-questions}.

\begin{figure}[t]
  \begin{center}
    \includegraphics[width=12cm]{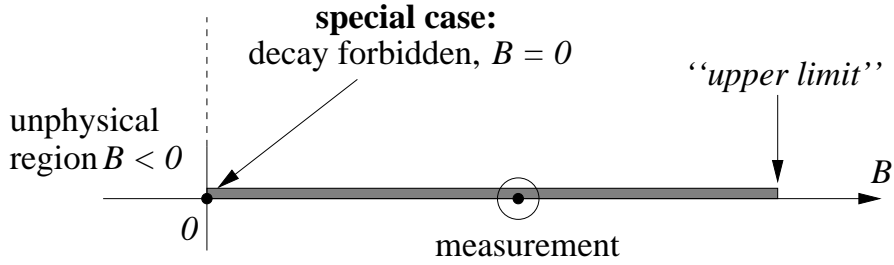}
  \end{center}
  \caption{A branching fraction measurement.
	The region $B<0$ is unphysical; $B=0$ is a special case.
	For a measurement $x = B_{meas}$,
	``integration of the likelihood function''
	(\emph{cf.}\ (\ref{eq-bayesian-90}))
	over the shaded region yields an upper limit as shown.}
  \label{figure-branching-flatprior}
\end{figure}

Having got that out of the way, I now want to explain why this method is not
as bad as it looks:
\begin{enumerate}
  \item	\emph{It is easy and fast, and can be done with information already
	``in hand''.} All you need is the likelihood function.
	This is important, since the typical rare decay analysis needs to be
	published with some urgency: either because it is a first observation,
	or because some issue hangs on the measurement.
	($\eta'\kmes(\pi)$ is of this kind: 
	there is a theory/data discrepancy which might be a first hint of  
	something exciting; see Table~\ref{table-etapk-tab2}.)
	There is a kind of built-in obsolescence in these measurements too,
	due to continuing improvements in our luminosity: Physics Letters
	published our paper~\cite{paper-etapk} on 4th October 2001: nine months
	after that date, we will have at least eight times its data on tape.
  \item	\emph{For branching fraction measurements, it corresponds roughly to
	a Bayesian interval.} If we have some
	prior degree of belief $p(\mu)$ concerning a parameter
	(here, $\mu = B_{true}$, the true branching fraction), then after the
	measurement $x$ we may update this belief using Bayes' Theorem
	\begin{equation}
	  p(\mu | x) = p(x | \mu) \cdot p(\mu) / p(x),
	  \label{eq-bayes}
	\end{equation}
	where $p(x | \mu) = \mathcal{L}(\mu ; x)$ is the likelihood function, 
	and $p(x)$ may be recovered from the normalization.	
	The posterior probability $p(\mu | x)$---our updated belief about $\mu$,
	following the measurement $x$---is a density in $\mu$ by construction,
	and therefore \emph{can} be integrated on $\mu$;
	and ``integrating the likelihood function'' is equivalent to
	integrating~(\ref{eq-bayes}) if the prior probability $p(\mu)$ is
	constant.
\end{enumerate}
Now a constant function is hard to defend as a serious prior, 
but there is a more subtle problem with this approach, to do with the special
point $\mu = 0$. If our prior $p(\mu)$ is truly constant, this means we are 
committed in advance to the belief that the branching fraction
\emph{does not vanish}, since $S = \{0\}$ is a set of measure zero: 
$\int_S \dcalc{\mu}\,p(\mu) = 0$ for any finite $p(\mu)$. If we were deriving
a proper Bayesian credible interval for $\mu$ we might well assign a
\emph{delta function} to the origin, allowing a (say) 10\% belief 
that the decay is forbidden; the posterior for $\mu = 0$ would then be
nonzero.\footnote{
	For $P(0 | x) = \int_S \dcalc{\mu}\,p(\mu | x) \ll 0.1$,
	the point $\mu = 0$ might lie \emph{outside} the 90\% interval,
	although na\"{\i}vely integrating $p(x | \mu)$ would never
	tell you that: (\ref{eq-bayesian-90}) always yields an 
	upper limit. That is, (\ref{eq-bayesian-90}) is not a 
	unified method for interval construction. 
	}
The upper limit calculated \emph{via}
\begin{equation}
    \textstyle
	\left[
  	{\int_0^B	\dcalc{\mu}\, p(x | \mu) \cdot p(\mu) / p(x)}
	\right]
    	\left/
	\left[
	{\int_0^\infty	\dcalc{\mu}\, p(x | \mu) \cdot p(\mu) / p(x)}
	\right]
	= 0.9
	\right.
  \label{eq-bayesian-90}
\end{equation}
would fall,
but this is just a special case of the dependence of the limit on the prior.
The special point $\mu=0$ is benign because it coincides with the physical
boundary (Fig.~\ref{figure-branching-flatprior}) and is always on the
\emph{edge} of an interval, if it belongs to it at all.
In section~\ref{section-pipi}, we will see an example where this is not the
case.

Upper limits derived in this way might differ by a factor of $\approx 2$
from those we would obtain by a more rigorous (Bayesian) technique.
Where there is already a tradition of quoting such limits,
so that everyone ``knows'' what they mean (just as we know that ``3 sigma'' and
``5 sigma'' do not \emph{really} mean 99.7\% and 99.994\% confidence),
it would be hard to justify declaring war on the method.
And the frequentist alternative is messy: it would require the construction
of toy Monte Carlos for each and every decay mode and analysis
(all of them differ subtly), to determine coverage \ldots
and we would probably find limits (again) within a factor 2, at the price 
of making lots of work (in parallel!) for lots of students.

%
%

\subsection{Fits over a ``flattish'' background}
\label{subsec-analyses-flattish}

These analyses are (to my mind) more straightforward: they involve fitting
a lineshape over a smooth background, or at worst, interpreting the result of
a background subtraction.
Our analysis of prompt charmonium production~\cite{paper-prompt-cc}
is an example. Selecting $J/\psi$ events with centre-of-mass momentum
$p^\ast_{\psi} > 2.0\,\gev/c$, above the kinematic limit for
$\bmes \to J/\psi\, X$, we measure the yield in the main ``on-resonance''
data, and in the smaller ``off-resonance'' sample, where $\sqrt{s}$ is 
just below $\bmes\bar{\bmes}$ threshold: too low to produce an \ufours\ meson.
After scaling, correction and cross-checks, we subtract the yields
to find the net number of $\ufours \to J/\psi\, X$ decays to be $-37 \pm 156$:
\emph{i.e.}\ consistent with vanishing $\mathcal{B}(\ufours \to J/\psi\,X)$.

The error is dominated by the uncertainty on the off-resonance
yield. We assume that this---and the full error---is 
distributed as a Gaussian, and use~\cite[table X]{feldman-cousins} to determine
the upper limit. This allows the negative yield to be treated in a rigorous way.
Some approximation is involved by assuming Gaussian behaviour,
and in principle we could model the subtraction in a toy Monte Carlo, 
and determine the limit using likelihood-ratio ordering from first
principles. In my view the utility of spending 30 seconds looking up a table, 
and referring readers to an accessible paper, outweighs issues of
principle here.  

\begin{figure}[t]
  \begin{center}
    \includegraphics[width=8cm]{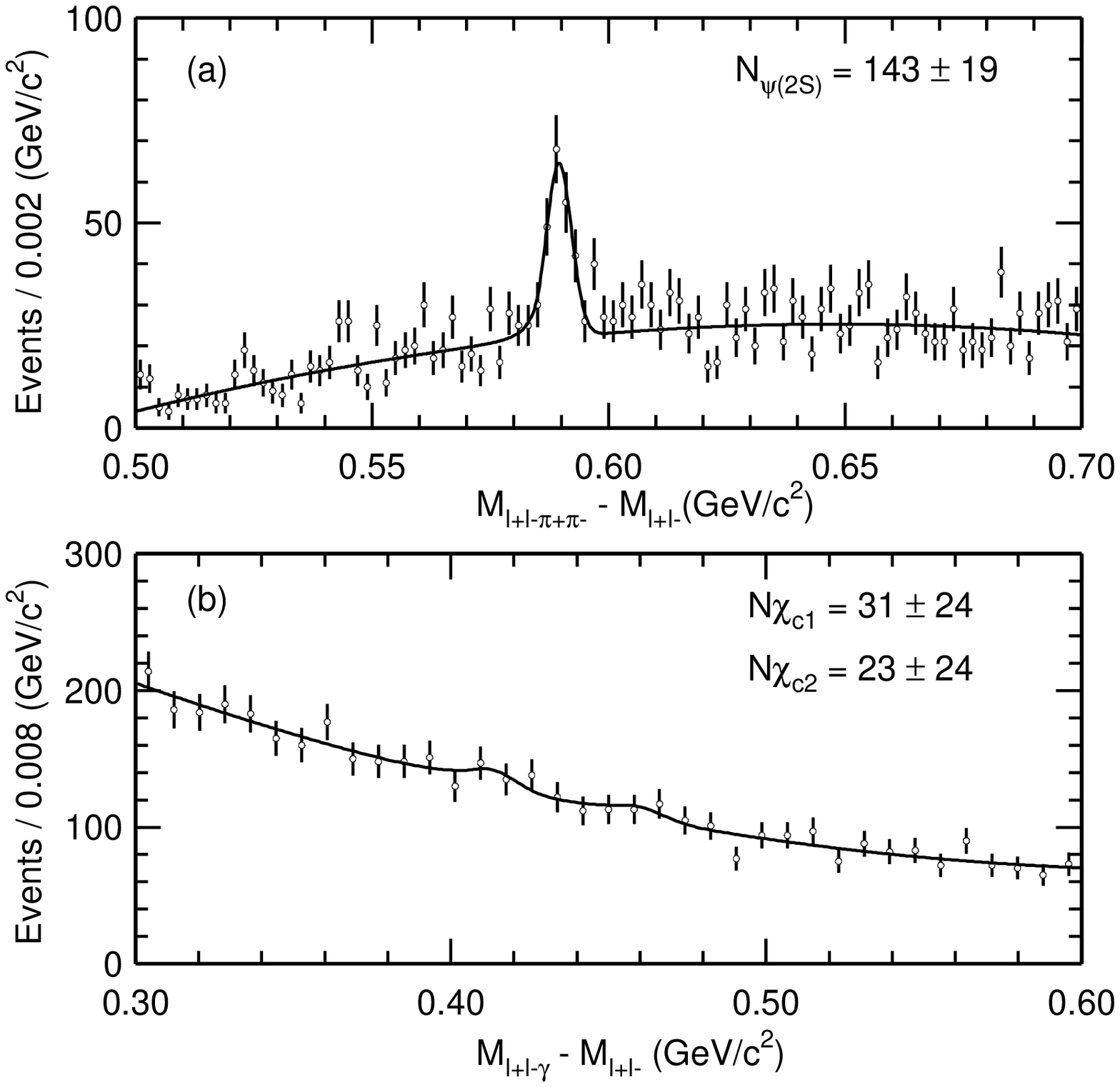}
  \end{center}
  \caption{From the prompt charmonium paper~\cite{paper-prompt-cc}:
	fits for the yield of $\psi(2S) \to J/\psi\,\pi^+\pi^-$ (upper plot)
	and $\chi_{c1,c2} \to \gamma J/\psi$ (lower plot) for
	$p^\ast > 2.0\,\gev/c$, above the limit for $\bmes \to (\ccbar)_{res}$
	decays.}	
  \label{figure-prompt-psiprime}
\end{figure}

Having set an upper limit for $\mathcal{B}(\ufours \to J/\psi\,X)$, we assume
that the observed prompt $J/\psi$ are produced directly from \epem\ 
annihilation. We look for prompt production of other charmonia, and find a
significant yield for $\psi(2S)$, but not $\chi_{c1,c2}$:
see Fig.~\ref{figure-prompt-psiprime}. For $\chi_{c1,c2}$ we set upper limits
on $\sigma(\epem \to \chi_{c1,c2}\,X)$ using the Feldman-Cousins tables for
Gaussians: the assumption of Gaussian errors is clearly reasonable.
The same technique is used when treating small yields on a background
in~\cite{paper-chic2}.

The search for (factorization-forbidden) $\bmes \to \chi_{c2}\,X$
decays~\cite{paper-chic2} is unusual for Belle: a \bmes-decay analysis where
we fit a peak over a large, smooth background, so statistical questions are
unproblematic. (Because of the inclusive nature of the decay, we cannot use 
the $M_{bc}, \Delta E$ variables to wipe away the background.)
There is however an interesting question concerning systematic errors. 
Because of the complicated (and overlapping) lineshapes used to fit
$\chi_{c1,c2} \to \gamma J/\psi$, and the relatively large $\chi_{c1}$ yield,
the systematic error on the $\chi_{c2}$ yield is substantial:
$\mathcal{B}(\bmes \to \chi_{c2}\,X) = (1.80^{+0.23}_{-0.28} \pm 0.26) \times 10^{-3}$.
The systematic error $\pm 0.26$ is dominated by the choice of the fit
function: we estimate the associated uncertainty using a large sample of 
reasonable (and some \emph{un}-reasonable) variations to the fitting model. 
How should we \emph{``combine''} the statistical and systematic errors in this
case? I, for one, don't know.

We \emph{can} answer the following,
more sharply posed question: is the yield ``significant'' even when
the systematics are taken into account?
(Phys. Rev. Lett. insists on ``$5\sigma$''
significance before you are allowed to call something an ``observation''.)
All attempted variations to the fit gave yields with significance $> 5\sigma$,
and we take this to be the relevant test: the statement that ``the yield is
inconsistent with fluctuations of the background'' does not depend on some
accidental feature of the fit, but is robust.

%
%

\subsection{Systematic-dominated analyses}
\label{subsec-analyses-systdominated}

When on the other hand systematic errors are dominant, we do not
quote intervals \ldots and questions of ``significance'' tend not to arise.
An example is our measurement of $\bmes \to X_s \gamma$
decays~\cite{paper-btosgamma}, where we find the underlying quark
transition---the theoretically interesting process---to have a branching
$\mathcal{B}(b\to s\gamma) =
(3.36 \pm 0.53\,\text{(stat.)}\; \pm 0.42\,\text{(syst.)}\;
		  ^{+0.50}_{-0.54}\,\text{(th.)}) \times 10^{-4}.$
There are plenty of issues of interpretation in such cases (they are 
\emph{analyses} properly-so-called) but they are beyond the scope of this
review.

%
%

\section{THE NEW $\bz/\bzbar \to \pi^+ \pi^-$ ANALYSIS}
\label{section-pipi}

As a final example, let's consider a very difficult problem: the interpretation
of our new $\bz/\bzbar \to \pi^+ \pi^-$ result~\cite{paper-pipi}.
Like the $\sin2\phi_1$ analysis, this is a measurement of a time-dependent
CP-violating asymmetry, with two differences: $\pi^+ \pi^-$
is sensitive to the unitarity angle $\phi_2$;
and there may be direct CP violation in the decay. We fit
\begin{equation}
  R_q(\Delta t) = \frac{e^{-|\Delta t|/\tau_B}}{4\tau_B}
		\left[ 1 + q \cdot \left\{
				A_{\pi\pi}\cos(\Delta m \Delta t) + 
				S_{\pi\pi}\sin(\Delta m \Delta t)
			\right\}
		\right]
  \label{eq-pipi-fit}
\end{equation}
to the decay-time distribution for \bz\ ($q=1$) and \bzbar\ ($q=-1$),
where $\tau_B$ and $\Delta m$ are the lifetime and mass-splitting of the 
eigenstates $|\bmes_\pm\rangle = p |\bz\rangle \pm q |\bzbar\rangle$.
The coefficients are given by
\begin{equation}
  A_{\pi\pi}	= \frac{|\lambda|^2 - 1}{|\lambda|^2 + 1},\;
  S_{\pi\pi}	= \frac{2\Im \lambda}{|\lambda|^2 + 1},\;
  \lambda	= \frac{q}{p}\cdot \frac{A(\bzbar \to \pi^+\pi^-)}
					{A(\bz    \to \pi^+\pi^-)}
  \label{def-pipi-coeff}
\end{equation}
A value $A_{\pi\pi} \neq 0$ corresponds to direct CP violation in the decay.
			
\begin{figure}[t]
  \begin{center}
    \includegraphics[width=12cm]{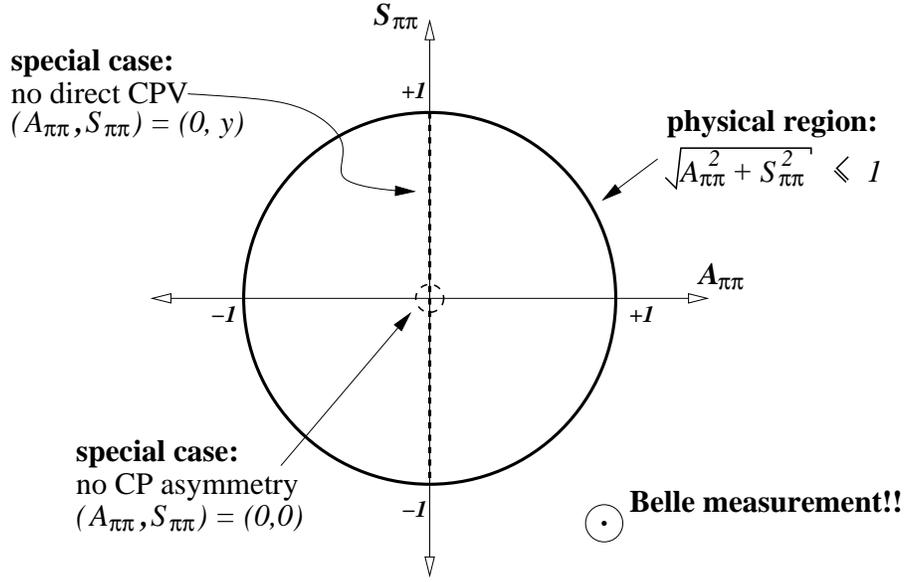}
  \end{center}
  \caption{The parameter space for the $\bz/\bzbar \to \pi^+\pi^-$ asymmetry
	analysis, with the Belle measurement.}
  \label{figure-btopipi}
\end{figure}

\begin{figure}[h]
  \begin{center}
    \includegraphics[width=8cm]{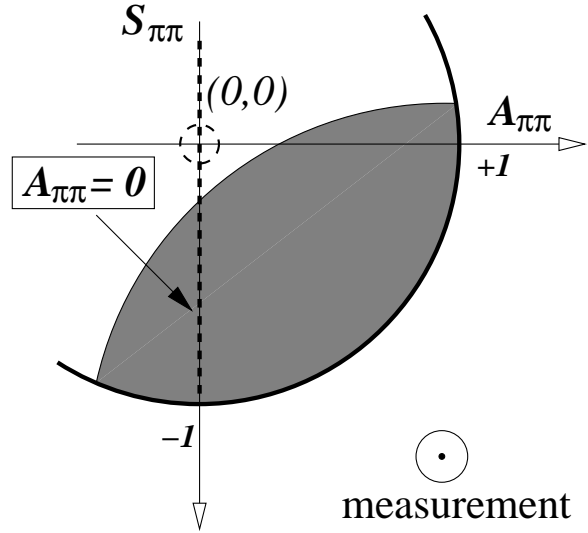}
  \end{center}
  \caption{Detail of Fig.~\ref{figure-btopipi}, showing an interval (shaded)
	constructed
	with a ``flat prior'' on the physical region. Note that the interval
	intersects the special region $A_{\pi\pi}=0$ before reaching the
	null asymmetry point $(0,0)$.} 
  \label{figure-btopipi-flatprior}
\end{figure}

The parameter space, Fig.~\ref{figure-btopipi},
is wonderfully complicated. There are two kinds of special region:
the null asymmetry point $(A_{\pi\pi},S_{\pi\pi}) = (0,0)$,
and the \emph{line} $(A_{\pi\pi},S_{\pi\pi}) = (0,y)$, where there is no
direct CPV ($A_{\pi\pi} = 0$).
The physical boundary $\sqrt{A_{\pi\pi}^2 + S_{\pi\pi}^2} = 1$ forms a ring 
around these regions \ldots and the Belle measurement
$A_{\pi\pi} = +0.94^{+0.25}_{-0.31} \pm 0.09$,
$S_{\pi\pi} = -1.21^{+0.25}_{-0.31}\, ^{+0.16}_{-0.13}$
lies \emph{outside it}!\footnote{Since the experimental quantity is an
	asymmetry, this is not possible for pure signal; \emph{cf.}\ a
	$\beta$-decay endpoint analysis, where resolution effects can give
	$m_\nu^2 < 0$. In the presence of background events, values outside
	the physical boundary can occur.}
If we want to determine a confidence interval in the true parameters
$(a_{\pi\pi},s_{\pi\pi})$, we immediately run into difficulty:
\begin{enumerate}
  \item	\emph{Frequentist method:}
	If we study the fit using Monte Carlo events, we find that the fitted
	error on $(A_{\pi\pi},S_{\pi\pi})$ \emph{varies} from virtual
	experiment to virtual experiment, by a factor of a few. So in 
	constructing a toy Monte Carlo to model the fit---and determine 
	confidence intervals---how do we generate the observed
	$(A_{\pi\pi},S_{\pi\pi})$, given some underlying parameters
	$(a_{\pi\pi},s_{\pi\pi})$?
	\vspace*{1ex}
	\begin{enumerate}
	  \item	If we use the \emph{measured} errors, what if we ``get lucky''
		and they are unusually small?
		Can we trust a result that says our value is ``inconsistent''
		with fluctuations from $(0,0)$?
	  \item	If we use the \emph{distribution} of errors from the full
		Monte Carlo, then the \emph{actual} errors returned by the fit
		are never used in the analysis.
		It seems paradoxical to ``throw away'' a fitted error.
	\end{enumerate}
	\vspace*{1ex}
  \item	\emph{Bayesian method:}
	Here the community expectation is less definite than for
	$\bz/\bzbar \to J/\psi\,\ks$, so any explicit prior would be
	controversial. But because of the configuration of the
	physically interesting points and the physical boundary, an explicit
	prior seems unavoidable.
	For suppose we tried to form a pseudo-Bayesian interval
	by ``integrating the likelihood function'' on the physical region,
	expanding outwards from the measured value
	(Fig.~\ref{figure-btopipi-flatprior}).
	We want to know if we reach the point of null asymmetry $(0,0)$ 
	before or after we have exhausted (say) 99.7\% of the integral:
	if the 99.7\% interval does not include $(0,0)$, surely we can say that
	we exclude it at some level \ldots and write off the improper treatment
	of the prior as a minor detail
	(\emph{cf.} section~\ref{subsub-flatprior})?

	This is not possible in general, because of the special region 
	$A_{\pi\pi} = 0$. If we allow some prior probability for
	indirect CPV ($S_{\pi\pi} \neq 0$) in the absence of direct CP
	violation, there is a ``delta-line function'' 
	$\delta(a) \cdot f(s)$ running along the $S_{\pi\pi}$ axis. 
	In general a credible interval will intersect this function
	\emph{before reaching $(0,0)$}. Thus there is no way to
	determine the probability content of the interval without making an
	explicit commitment as to the prior $\delta(a) \cdot f(s)$.
	(A ``flat'' prior means $f(s)=0$.)
	I can see no way around this problem: any prior must be explicit.
\end{enumerate}

%
%

\section{SOME OPEN QUESTIONS}
\label{section-questions}

\subsection{How do we quote rare decay results?}

In section~\ref{subsec-analyses-rare}, in all the excitement about priors, 
I almost forgot to notice a glaring inconsistency.
We require significance $\Sigma > 3$ before we will quote a 
central value, corresponding to a 99.7\% confidence requirement \ldots
but we quote 90\% C.L. upper limits! If a value falls between 
these two levels, we should not be able to say anything at all: in fact we are
let off the hook because our interval-setting is not unified.
(As noted above, the integral method~(\ref{eq-bayesian-90}) always gives an
upper limit.)
In defence of Belle, we are following community practice and expectations
here, and those expectations are incoherent. My provisional idea for a way
around the problem is to
\begin{itemize}
  \item	\emph{always} quote the central value;
  \item	construct 99.7\% C.L. intervals in a unified manner (\emph{i.e.}\ 
	going over continuously from upper limits to central intervals, the
	way the so-called Feldman-Cousins intervals~\cite{feldman-cousins} do);
  \item	use these intervals in place of the $\Sigma > 3$ test;
  \item	quote 90\% C.L. intervals \emph{as well}, because people expect them;
  \item	if people query the use of 3 numbers/intervals, rather than 1, explain;
  \item	if people \emph{object} to the use of 3 numbers, resort to violence.
\end{itemize}
	
This would be consistent, but it is a utopian scheme \ldots and I suspect it
would take a dictator to implement it
(\emph{cf.}\ section~\ref{section-practice}).
I will try to raise consciousness on this issue at Belle, but it really is a 
community problem, and we should try to think up some way for all of us to get
to ``there'' from ``here''. The three-values approach I've suggested may not
be the best way.

Note also that a unified treatment of ``significance'' and 
confidence intervals begs another question:

\subsection{How do we combine statistical and systematic errors?}

In sections~\ref{subsec-analyses-timing} and~\ref{subsec-analyses-flattish}
I noted cases where we are able to avoid this question, but this is not general.
The technique used for the rare decays (section~\ref{subsec-analyses-rare})
is peculiar: the integration method is pseudo-Bayesian, as discussed at some
length; it may not be quite so obvious, but the practice of inflating the
confidence intervals by ``one sigma'' of the systematic error is 
pseudo-Frequentist.
It treats the systematic error as a nuisance
parameter with range $[-\delta^-, +\delta^+]$, and demands that our
confidence interval provides coverage for all values in that region.
To my mind this is almost exactly the wrong way around:
\begin{itemize}
  \item	I think our prior beliefs about branching fractions and CP violation
	are not of interest (or at least, do not belong in papers), which 
	suggests a Frequentist approach; whereas
  \item	our beliefs about our systematic errors surely are relevant---everyone
	knows they come down to a question of judgement---which suggests a
	Bayesian approach.
\end{itemize}
A complicating factor is that not all ``systematic errors'' are the same kind
of thing:
\begin{itemize}
  \item	Particle ID efficiencies are measured on control samples, which have
	statistical uncertainties: it seems reasonable to use some averaging
	method when assessing the effect on any final interval.
  \item	The unknown phases of resonances on Dalitz plots (for example) are at
	the other extreme: we have no business making assumptions about them, 
	and if they significantly affect a result we should require any 
	confidence interval to provide coverage for all possible values.
  \item	The choice of parameterization for a fit function
	is unlike both of the previous examples.
\end{itemize}
Needless to say, these musings do not constitute a policy, much less a recipe.
I suspect that intervals of the form
$(x - 1.64\sigma^- - \delta^-,\; x + 1.64\sigma^+ + \delta^+)$ will remain
with us for some time.

\subsection{What should we do about the $\pi^+\pi^-$ analysis?}

The consensus from the floor during this talk was that in a problem of this 
difficulty, rigour is essential: only the two extreme approaches make sense,
\emph{viz.}
\begin{enumerate}
  \item	a Frequentist calculation from first principles, and
  \item	a full, openly subjective Bayesian analysis.
\end{enumerate}
Regarding the second of these, I am already convinced
(see section~\ref{section-pipi}). As for the Frequentist calculation, I am 
indebted to my colleagues at the conference for some ingenious suggestions
on the proper treatment of the errors.
I hope to experiment with them, together with my Belle colleagues, before the
summer.

%
%

\section{CONCLUSION}
\label{section-conclusion}

The Belle collaboration performs a large range of analyses, using a range of
statistical approaches. Some of the methods are open to question, 
although there is a clear trade-off between utility and statistical rigour
in some cases. The expectations of the broader physics community also play
a role.
In the case of the $\bz/\bzbar \to \pi^+\pi^-$ analysis,
the statistical environment is unusually difficult, and rigorous (rather than
approximate) methods are required.

\vskip1cm
\noindent

\section*{ACKNOWLEDGEMENTS}

I would like to thank the conference organizers for arranging an instructive
and entertaining programme for this meeting. In particular, I am indebted to
the conference secretary Linda Wilkinson, for her help in recovering lost
slides and other material.
To my colleagues on Belle, and the collaboration management: thanks for their
patience with my questions and criticisms on statistical matters,
and for the freedom I am habitually given to speak my mind in public.

\end{document}